# The Width of the Colour Flux Tube

M. Caselle[a], F. Gliozzi[a]*, U. Magnea[b] and S. Vinti[a]

[a]Dip. di Fisica Teorica dell'Università and INFN, Sezione di Torino, I-10125 Torino, Italy

[b]State University of N.Y. at Stony Brook, Stony Brook, NY 11794 USA

We discuss and rederive in a general way the logarithmic growth of the mean squared width of the colour flux tube as a function of the interquark separation. Recent data on 3D $\mathbb{Z}_2$ gauge theory, combined with high precision data on the interface physics of the 3D Ising model fit nicely this behaviour over a range of more than two orders of magnitude.

## 1. THE FLUX TUBE AS A STRING

One of the simplest and most general consequences of the existence of a thin flux tube connecting a quark pair in the confining phase is that it fluctuates. These quantum fluctuations produce an effective squared width of the flux tube which grows logarithmically with the interquark distance.

Such a logarithmic behaviour has been predicted many years ago by Lüscher, Münster and Weisz [1] in the framework of the effective string picture of gauge systems. We shall rederive and refine this universal law by directly using some exact results on two dimensional free field theory in a finite box and compare it with MC simulations on the 3D $\mathbb{Z}_2$ gauge model.

Let us begin by defining the observable we want to discuss. The squared width of the flux tube generated by a planar Wilson loop $W(C)$ is defined as the sum of the mean squared deviations of the transverse coordinates $h(\xi_1, \xi_2)$ of the underlying string, i. e.

$$w^2 = \frac{1}{A} \sum_{i=1}^{D-2} \int_{\mathcal{D}} d^2\xi \langle \left(h_i(\xi_1, \xi_2) - h_i^{CM}\right)^2 \rangle \quad , \quad (1)$$

where $\mathcal{D}$ is the planar domain bounded by $C$, $A$ its area $A = \int_{\mathcal{D}} d^2\xi$ and $h_i^{CM}$ is the transverse coordinate of the center of mass of the flux tube.

---
*Speaker at the conference

In terms of the Green functions $G_i$ we get

$$w^2 = \frac{1}{\sigma A^2} \sum_{i=1}^{D-2} \int_{\mathcal{D}} d^2\xi \int_{\mathcal{D}} d^2\xi' F_i(\xi, \xi') \quad (2)$$

with

$$F_i(\xi, \xi') = G_i(\xi, \xi') - G_i(\xi, \xi + \varepsilon) \quad , \quad (3)$$

where $\varepsilon$ is a UV cut-off, $\sigma$ is the string tension and the Green function is defined as

$$G_i(\xi, \xi') = \frac{1}{\sigma} \langle h_i(\xi) \, h_i(\xi') \rangle \quad . \quad (4)$$

The vacuum expectation value is taken with respect to the 2D field theory describing the dynamics of the flux tube (for a review on the effective string description, see Ref. [2]). It is widely believed that for $\beta$ larger than the roughening point the flux tube belongs to the rough phase of the Kosterlitz - Thouless universality class. As a consequence it is expected that this field theory flows, for large enough domain $\mathcal{D}$, to the gaussian limit, where the Green function fulfills the free field equation $-\Delta G(\xi, \xi') = \delta^{(2)}(\xi - \xi')$ .

The problem of finding the Green function for an arbitrary, simply connected region $\mathcal{D}$ can be solved in a closed form once it is found a conformal mapping $z = \xi_1 + i\xi_2 \to \ell$ of $\mathcal{D}$ onto the unit circle $|\ell| = 1$ which maps $z' = \xi'_1 + i\xi'_2$ into the origin $\ell = 0$. Denoting by $\ell_{z'}(z)$ the analytic function providing us with such a mapping, it is immediate to verify that the real function $f_{z'}(z, \mathcal{D}) = \log|\ell_{z'}(z)|$ is harmonic in the punctured set $\mathcal{D} \setminus \{z'\}$ , vanishes at the boundary $\partial \mathcal{D}$

and diverges logarithmically as $-\log|z-z'|$ for $z \to z'$. It follows that the Green function is given by

$$G_{\mathcal{D}}(z,z') = \frac{1}{2\pi} f_{z'}(z,\mathcal{D}) \quad . \quad (5)$$

For our purposes, the relevant property of the conformal mapping $\ell$ is that one can perform an arbitrary scale transformation $z \to \Lambda z$ without destroying the conformal character of $\ell$. More precisely we can write

$$f_{z'}(z,\mathcal{D}) = f_{\Lambda z'}(\Lambda z, \mathcal{D}_\Lambda) \quad , \quad (6)$$

where $\mathcal{D}_\Lambda$ denotes the scaled domain. A direct consequence is that it is always possible to fix the area of the scaled domain $\mathcal{D}_\Lambda$ to an arbitrary value, say 1, without changing the Green function. It follows that the integration of the finite part $G(z,z')$ in Eq.(2) cannot depend on the size of the domain $\mathcal{D}$ but only on its shape. The logarithmic growth of the squared width $w^2$ comes from the UV divergent part. Combining the logarithmic divergence of $G$ with its scaling property expressed in Eq.(6), we can write

$$G(\xi', \xi' + \varepsilon) \sim -\frac{1}{2\pi} \log(\varepsilon/L) + c \quad , \quad (7)$$

where $L$ is a typical linear dimension of the domain $\mathcal{D}$. Inserting this expression in Eq.(2) we get the logarithmic law

$$w^2 = \frac{1}{2\pi\sigma} \log(L/R_c) \quad , \quad (8)$$

where the UV cut-off has been absorbed in the definition of the scale $R_c$, which is a calculable function of the shape of the domain.

The physical meaning of such a behaviour is now clear: the dynamics of the flux tube is described by a truly free field theory only at large distances; the cut-off $\varepsilon$ sets up the ultraviolet scale $R_c$ below which the free-field approximation breaks down. Obviously this scale cannot depend on the infrared scale $L$, then a variation of $L$ cannot be balanced by a variation of $\varepsilon$, as the scale invariance of Eq.(6) would require.

For a rectangular box of size $L_1 \times L_2$, which is the most interesting case for the lattice gauge models, we have

$$G_{L_1,L_2}(z,z') = -\frac{1}{2\pi} \log \left| \frac{\sigma(z-z')\sigma(z+z')}{\sigma(z-\bar{z}')\sigma(z+\bar{z}')} \right| \quad (9)$$

where $\sigma(z)$ is the Weierstrass sigma function for the rectangle of sides $2L_1$ and $2L_2$. Inserting this expression in Eq.(2) it turns out that the scale $R_c$ does not depend very much on the ratio $L_2/L_1$ for not too elongated rectangles.

## 2. NUMERICAL SIMULATIONS

Though the logarithmic growth of the squared width of the flux tube is the most important and model-independent quantum effect predicted by the effective string description, it has not yet been observed until now because it is an infrared phenomenon that can be seen only in very large Wilson loops, which are at the limit of the sizes reached by present numerical simulations on $SU(2)$ and $SU(3)$ 4D gauge theories.

It is nowadays possible to overcome this problem in the $\mathbb{Z}_2$ 3D gauge model by exploiting its duality relationship with the ordinary 3D Ising model. Indeed, using the one-to-one mapping of physical observables of the gauge system into the corresponding spin observables, it is possible to replace the ordinary Metropolis or heath-bath method with a much more efficient non-local cluster algorithm (in our case the Swendsen-Wang method). This allows us to probe the structure of the flux tube with an unprecedented accuracy. The procedure is the following. A Wilson loop $W(C)$ is realized in the spin lattice by frustrating all the links cutting a given surface $\Sigma$ bounded by $C$. These frustrated links modify the vacuum state so that the expectation value $\langle P \rangle_W$ of the plaquette, or better its spin counterpart, becomes a function of its relative position with respect to $W(L_1, L_2)$ and is related to the expectation value in the ordinary vacuum by

$$\langle P \rangle_W = \langle W(L_1, L_2) P \rangle / \langle W(L_1, L_2) \rangle \quad . \quad (10)$$

The difference between the expectation value of the plaquette in the vacuum modified by the presence of $W(L_1, L_2)$ and in the ordinary vacuum can be considered as a measure of the density of the flux tube. Choosing for instance as a probe a plaquette $P_\parallel$ parallel to the plane of the Wilson loop we can take

$$\rho_\parallel(x,y,z) = \langle P_\parallel \rangle_W - \langle P \rangle \quad . \quad (11)$$

Other orientations of the plaquette give approximately the same distribution [†]. Assuming that the Wilson loop is located in the plane $z = 0$, the mean squared width of the flux tube can be defined as

$$w^2 = \int dx\, dy\, dz\, z^2 \rho_\parallel \Big/ \int dx\, dy\, dz\, \rho_\parallel \ . \qquad (12)$$

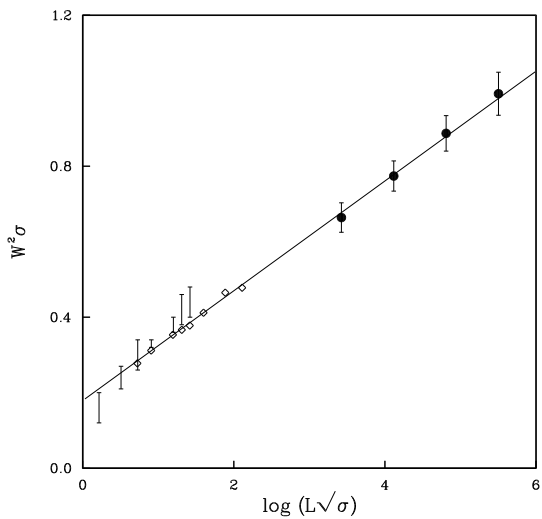

Figure. 1. Squared width of the $\mathbb{Z}_2$ flux tube as a function of interquark distance.

We performed two sets of simulations [6]. The first was done at $\beta = 0.7516$ corresponding to $\sigma a^2 = 0.0107(1)$. We considered six different square Wilson loops of sides ranging from $L = 16$ to $L = 40$ and fitted the data to the two-parameter formula

$$w^2 = a\, \log L + b \ . \qquad (13)$$

We got $b = -17.4 \pm 4.0$ and $a = 14.4 \pm 1.2$. The theoretical value $a_{th}$ of the parameter $a$, fixed by Eq.(8) to be $a_{th} = \frac{1}{2\pi\sigma}$, matches nicely with this value, indeed we have $a_{th} = 14.8$. The second set of simulations was done at $\beta = 0.7460$ corresponding to $\sigma a^2 = 0.0189(1)$ with square Wilson loops of sides ranging from $L = 15$ to $L = 60$. Fits to Eq.(13) gave $b = -5.8 \pm 2.0$ and $a = 7.7 \pm 0.6$ while the theoretical value is $a_{th} = 8.42$.

An internal consistency check of these numerical data comes from a comparison between Eq.(8)

---
[†]This seems not to be the case in four dimensional systems[3,4]

and Eq.(13). We get that the physical adimensional quantity $\sqrt{\sigma} R_c$ is expressed in terms of $a$ and $b$ as

$$\sqrt{\sigma} R_c = \frac{\exp -b/a}{\sqrt{2\pi a}} \ . \qquad (14)$$

Using the fitted values of the parameters we get $\sqrt{\sigma} R_c = 0.35 \pm 0.11$ at $\beta = 0.7516$ and $\sqrt{\sigma} R_c = 0.31 \pm 0.09$ at $\beta = 0.7560$ in good agreement with scaling.

One important feature of Eq.(8) is that it can be written in a universal form by expressing all the dimensional quantities $w$, $L$ and $R_c$ in units of $\sqrt{\sigma}$. Accordingly we report in Fig.1 data from different $\beta$'s and also very accurate recent data[5] for the mean squared width of fluid interfaces in the dual Ising model (black dots). The straight line represents a logarithmic fit to our data at $\beta = 0.7460$ (rhombs). Within the statistical accuracy these data clearly support a logarithmic widening of the flux tube in a range of quark separation $L$ over more than two orders of magnitude, starting at about $\sqrt{\sigma} L_{min} \simeq 1.7$.

In order to compare these results with analogous data for other gauge groups, we may, with an abuse of language, express $\sqrt{\sigma}$ in same physical units as the QCD string. Then the crossover to the infrared logarithmic behaviour observed in Fig. 1 corresponds to $L_{min} \simeq 0.75$ fm while the maximal probed elongation corresponds to more than 100 fm. The data on 4D $SU(2)$ flux tube[3,4] cover now a distance up to 2 fm, but are still affected by strong systematic errors and are compatible also with a constant width for distances larger than 1 fm.